\documentclass[aps,preprint,12pt]{revtex4}
\usepackage{amsfonts}
\usepackage{amsmath}
\usepackage{amssymb}
\usepackage{graphicx}
\usepackage{dcolumn}
\providecommand{\U}[1]{\protect\rule{.1in}{.1in}}
\newcommand{\bfr}{{\bf r}}
\newcommand{\bfR}{{\bf R}}
\newcommand{\up}{\uparrow}
\newcommand{\dn}{\downarrow}
\newcommand{\updn}{{+}{-}}
\newcommand{\bfq}{{\bf q}}
\newcommand{\bfk}{{\bf k}}
\begin{document}
\title[CaFe$_{2}$As$_{2}$]{Low Energy, Coherent, Stoner-like Excitations in CaFe$_{2}$As$_{2}$}
\author{Liqin Ke$^{1}$, M. van Schilfgaarde$^{1}$, J.J.Pulikkotil$^{2}$, T.
Kotani$^{3}$, and V.P.Antropov$^{2}$}
\affiliation{$^{1}$ School of Materials, Arizona State University}
\affiliation{$^{2}$ Ames Laboratory, IA 50011}
\affiliation{$^{3}$ Tottori University, Tottori, Japan}
\pacs{PACS number}

\begin{abstract}

Using linear-response density-functional theory, magnetic excitations in
the striped phase of CaFe$_{2}$As$_{2}$ are studied as a function of local
moment amplitude.  We find a new kind of excitation: sharp resonances of
Stoner-like (itinerant) excitations at energies comparable to the
N{\'{e}}el temperature, originating largely from a narrow band of Fe $d$
states near the Fermi level, and coexist with more conventional (localized)
spin waves.  Both kinds of excitations can show multiple branches,
highlighting the inadequacy of a description based on a localized spin model.

\end{abstract}
\eid{identifier}
\date{\today}
\maketitle

%\affiliation{$^{3}$ Tottori University, Tottori, Japan}

Magnetic interactions are likely to play a key role in mediating
superconductivity in the recently discovered family of iron pnictides; yet
their character is not yet well understood.  In particular, whether the
system is best described in terms of large, local magnetic moments centered
at each Fe site, in which case elementary excitations are collective spin
waves (SWs) called magnons, or is itinerant (elementary excitations characterized
by single particle electron-hole transitions) is a subject of great debate.
This classification also depends on the energy scale of interest.  The most
relevant energy scale in CaFe$_{2}$As$_{2}$ and other pnictides ranges to
about twice the N{\'{e}}el temperature, 2$T_{N}${}$\approx$40 meV.
Unfortunately, neutron scattering experiments have focused on the character
of excitations in the 150-200 meV range \cite{AMESneut,ORNLneut}, much
larger than energy that stabilizes observed magnetism or superconductivity.
Experiments in Refs.~\cite{AMESneut,ORNLneut}, while very similar, take
completely different points of view concerning the magnetic excitations
they observe.  There is a similar dichotomy in theoretical analyses of
magnetic interactions\cite{Mazin1,JIJ}.  Model descriptions usually
postulate a local-moments picture~\cite{LOCMOD}.  Most \emph{ab initio}
studies start from the local spin-density approximation (LSDA) to density
functional theory.  While the LSDA traditionally favors itinerant
magnetism (weak on-site Coulomb correlations), practitioners strongly
disagree about the character of pnictides; indeed the same results are used
as a proof of both localized and itinerant descriptions~\cite{JIJ,Mazin1}.

The dynamic magnetic susceptibility (DMS), is the central quantity that
uniquely characterizes magnetic excitations. It can elucidate the origins
of magnetic interactions and distinguish between localized and itinerant
character.  However, the dynamical linear response is very difficult to
carry out computationally; studies to date been limited to a few simple
systems.  Here we adapt an all-electron linear response technique developed
recently \cite{GWMETHOD,SWMNO} to calculate the transverse DMS,
$\chi(\mathbf{q},\omega)$.  Besides SW excitations seen in neutron
scattering, we find low energy particle-hole excitations at low $q$, and
also at high $q$.  In stark contrast to conventional particle-hole
excitations in the Stoner continuum, they can be sharply peaked in energy
(resonances) and can be measured.

%Make out.ps with
%catps -o out.ps dosef.eps -t 0 4 in dos.wideE.eps ; cat out.ps | grep -v BoundingBox  >out.eps; fixboundingbox out.eps
%\begin{center}
\begin{figure*}[ptb]
\includegraphics[height=3.3cm]{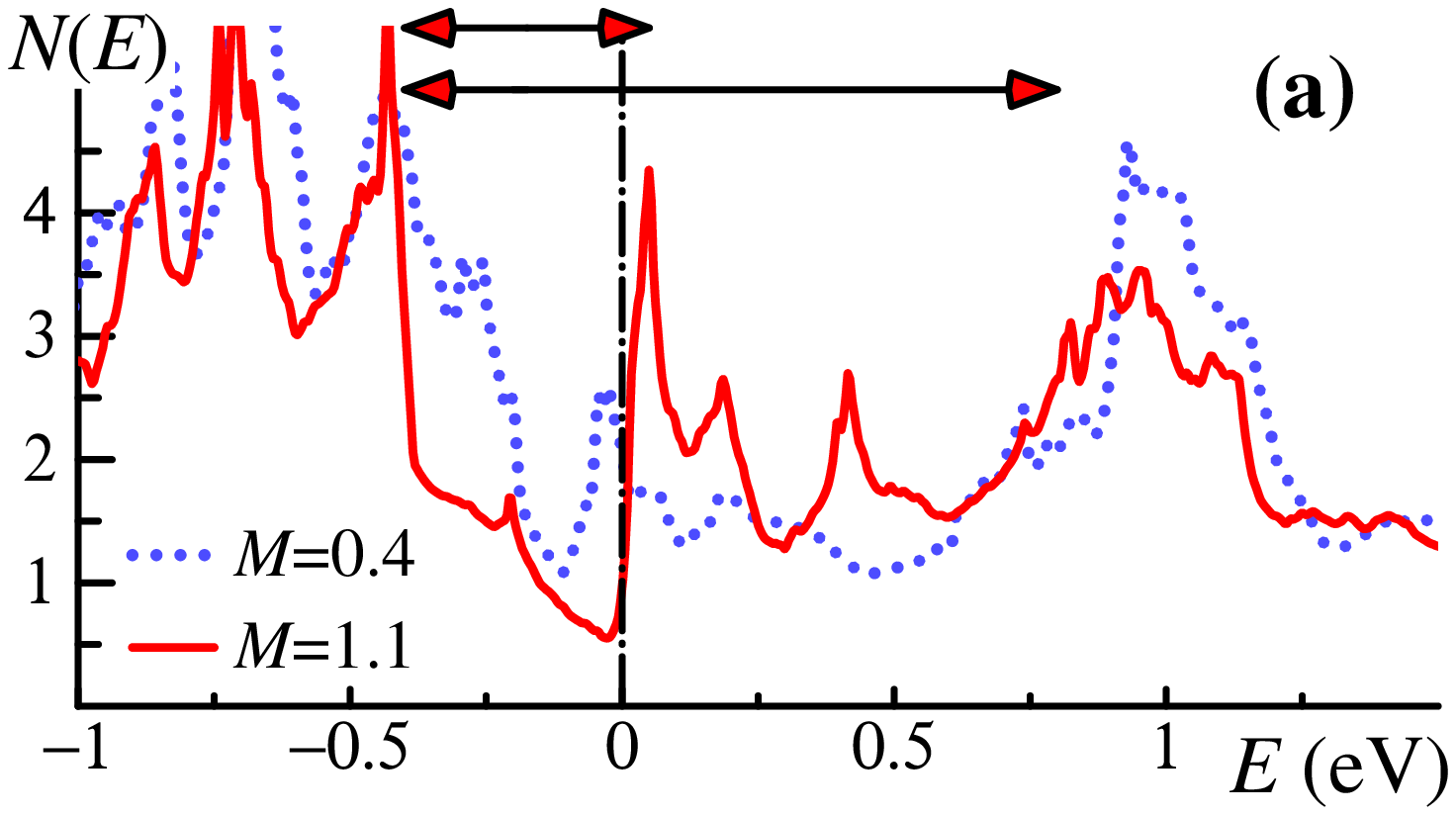} \
\includegraphics[height=3.3cm]{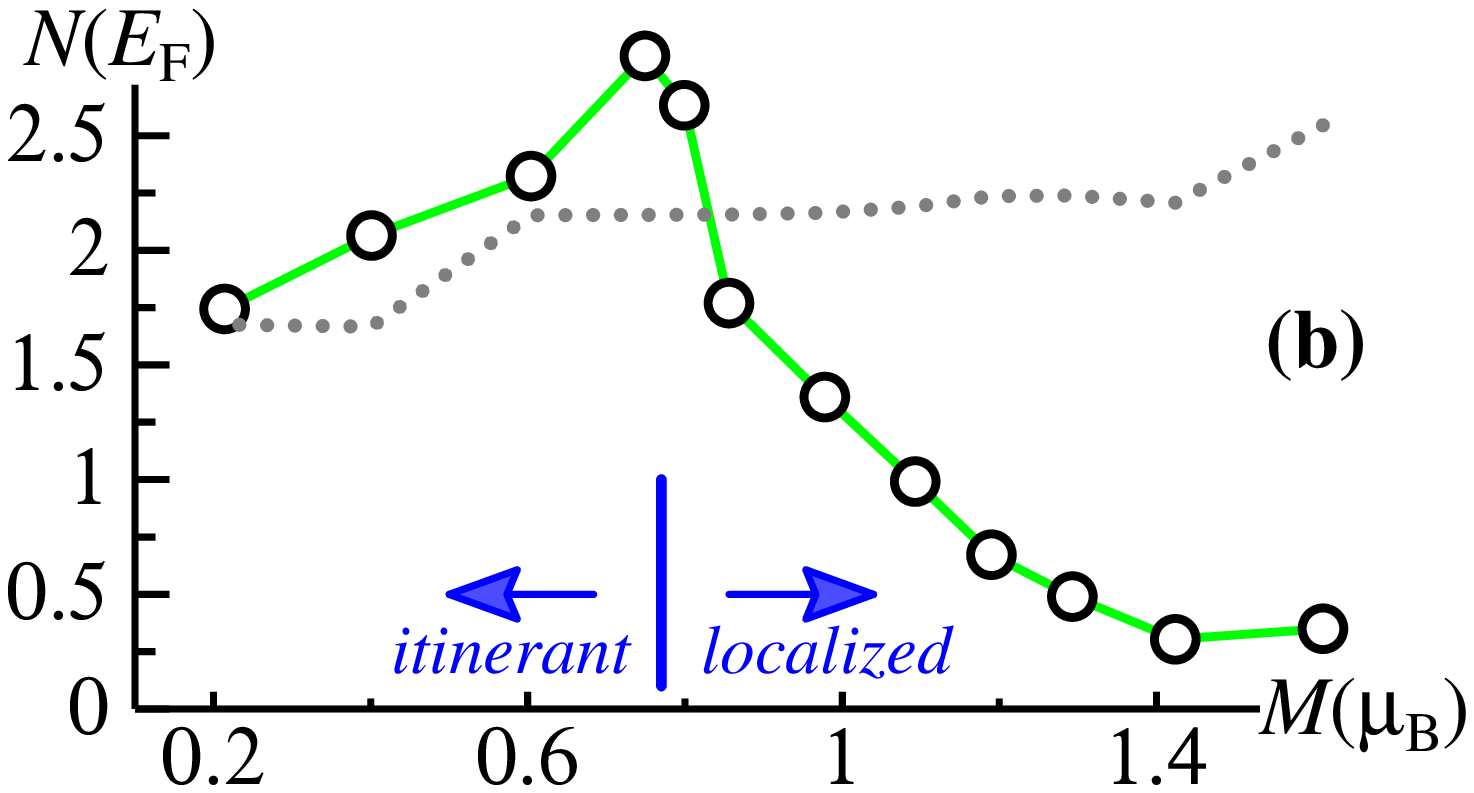}
\includegraphics[height=3.3cm]{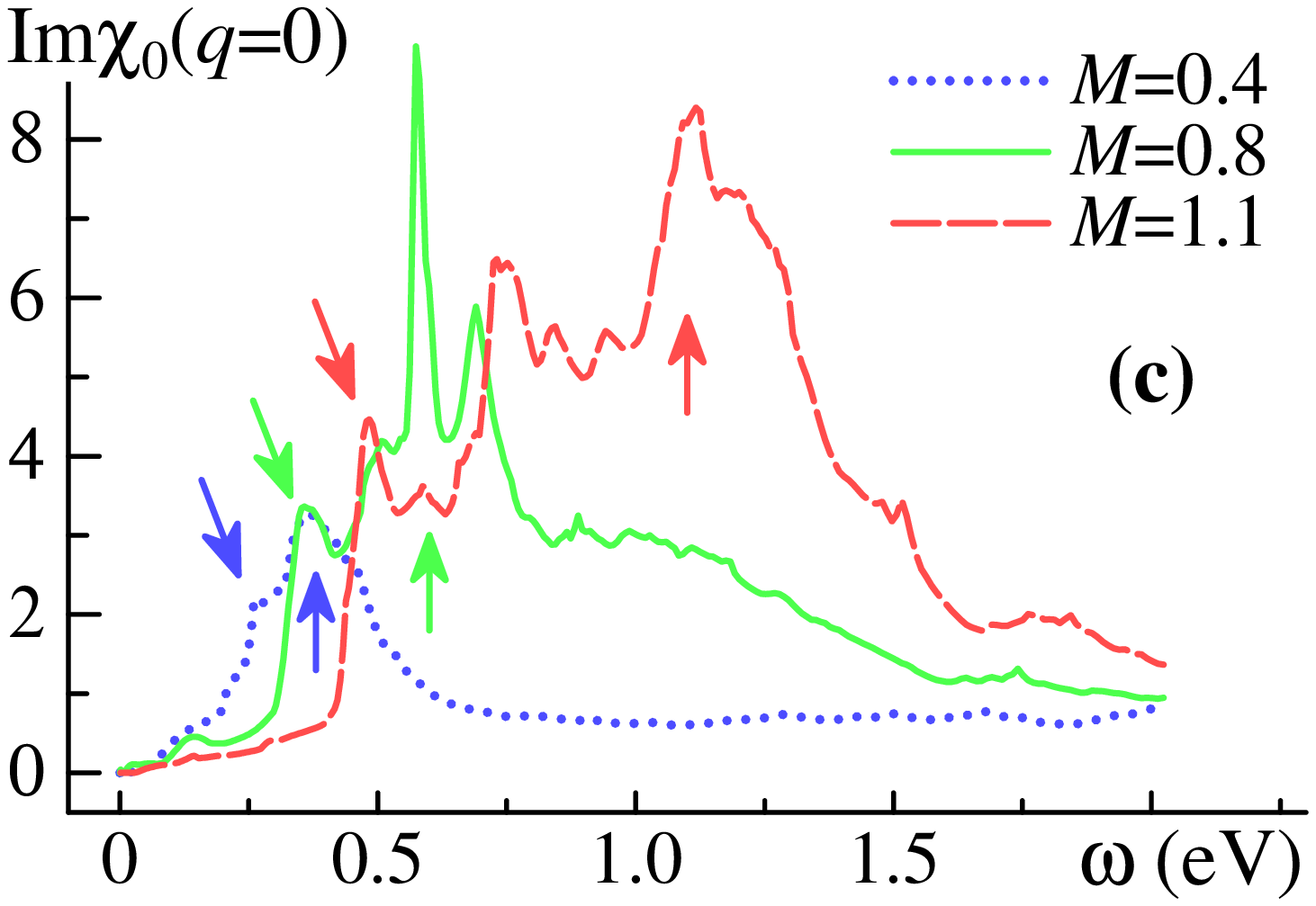}
%\includegraphics[width=5.0cm]{chi0q0.eps}
%\includegraphics[angle=0,width=0.50\textwidth,clip]{dos.wideE.eps}
%\includegraphics[width=5.4cm]{out.eps} \
%\begin{tabular}{cc}
%\includegraphics[height=2.8cm]{doskefm10meV.eps} &
%\includegraphics[height=2.8cm]{doskefp10meV.eps}
%\end{tabular} \\
%\includegraphics[width=5.4cm]{k-dos.eps} \ \
%\includegraphics[width=5.4cm]{chi0q0.eps}
\caption{(\textbf{{a}}) $N(E)$, in units of eV$^{-1}$ per cell containing
one Fe atom. Data are shown for $M$=0.4$\mu_{B}$ and 1.1$\mu_{B}$.  Both kinds of transitions are
reflected in peaks in $\chi_{0}$($q$=0,$\,\omega$), shown in panel
(\textbf{{c}}) for $M$=0.4, 0.8, and 1.1$\mu_{B}$. (\textbf{{b}}) $N(E)$ at
the Fermi level $E_{F}$ as a function of moment $M$. The blue vertical bar
denotes the experimental moment, and also approximately demarcates the
transition from itinerant to localized behavior.  $N(E_{F})$ in the
nonmagnetic case is shown as a dotted line. (\textbf{{c}}) bare
susceptibility $\chi_{0}$($q$=0,$\,\omega$) in the same units.
The text discusses the significance of the arrows in panels (\textbf{{a}})
and (\textbf{{c}}).
}
\label{fig:dos}
\end{figure*}
%\end{center}
%\includegraphics[angle=0,width=0.50\textwidth,clip]{chi0q0.eps}

%--- Formal development --
The full transverse DMS
$\chi(\mathbf{r},\mathbf{r}^{\prime},\mathbf{q},\omega)$ is a function of
coordinates $\mathbf{r}$ and $\mathbf{r}^{\prime}$ (confined to the unit
cell).  It is obtained from the non-interacting susceptibility
$\chi_{0}(\mathbf{r},\mathbf{r}^{\prime},\mathbf{q},\omega)$ via the
standard relation~\cite{Martinbook}
\begin{equation}
\chi=\chi_{0}\left[  1-\chi_{0}I\right]  ^{-1}
\label{eq:chi}
\end{equation}
$I$ is the exchange-correlation kernel. When computed within the
time-dependent LSDA (TDLDA) $I$ is local: $I$=$I(\mathbf{r})\delta
(\mathbf{r}-\mathbf{r}^{\prime}$)~\cite{Martinbook}. $\chi_{0}$ can be
obtained from the band structure using the all-electron methodology we
developed~\cite{GWMETHOD,SWMNO}.  To reduce the computational
cost we calculate Im$\chi_{0}$ and obtain Re$\chi_{0}$ from the Kramers Kronig
transformation~\cite{GWMETHOD}. Im$\chi_{0}$ originates from spin-flip
transitions between occupied states at $\mathbf{k}$ and unoccupied states at
$\mathbf{k}+\mathbf{q}$: it is a $k$-resolved joint density of states $D$
decorated by products $P$ of four wave functions~\cite{SWMNO}
\begin{eqnarray}
D(\bfk,\bfq,\omega) =
f(\epsilon^{\up}_{\bfk})(1-f(\epsilon^{\dn}_{\bfq+\bfk}))
\delta(\omega-\epsilon^{\dn}_{\bfq+\bfk}+\epsilon^{\up}_{\bfk})
\label{eq:jdos} \\
{\rm{Im}}\chi_0^{\updn} =
\int d\omega d^3\bfk P(\bfr,\bfr',\bfk,\bfq) \times
D(\bfk,\bfq,\omega)
\label{eq:chi0}
\end{eqnarray}
Even with the Kramers-Kronig transformation, Eq.~(\ref{eq:chi0}) poses a
huge computational burden for the fine frequency and $k$ resolution
required here.  We make a simplification, mapping $\chi_{0}$ onto the local
magnetization density which is assumed to rotate rigidly. The full
$\chi_{0}(\mathbf{r},\mathbf{r}^{\prime},\mathbf{q},\omega)$ simplifies to
the discrete matrix
$\chi_{0}(\mathbf{R},\mathbf{R}^{\prime},\mathbf{q},\omega)$ associated
with pairs of magnetic sites $(\mathbf{R},\mathbf{R}^{\prime})$ in the unit
cell; and $I(\mathbf{r})$ simplifies to a diagonal matrix
$I_{\mathbf{R}\mathbf{R}}$. In Ref.~\cite{SWMNO} we show that we need not
compute $I$ explicitly but can determine it from a sum rule.  $I$ can be
identified with the Stoner parameter in models. We have found that for Fe
and Ni these approximations yield results in rather good agreement with the
full TDLDA results, and expect similar agreement here. We essentially
follow the procedure described in detail in Ref.~\cite{SWMNO}, and obtain
$\chi(\mathbf{q},\omega)$ as a $4\times{}4$ matrix corresponding to the
four Fe sites in the unit cell.  To make connection with neutron
experiments, spectra are obtained from the matrix element
$\sum_{{\mathbf{R},\mathbf{R}^{\prime}}}<e^{i\bfq\cdot\bfR}|
\chi({\mathbf{R},\mathbf{R}^{\prime}},\mathbf{q},\omega) |
e^{i\bfq\cdot\bfR^\prime}>$.  For brevity we omit indices
${\mathbf{R}\mathbf{R}^{\prime}}$ henceforth.

%--- Ground state properties ---

As we will see, the character of $\chi_{0}(\mathbf{q},\omega)$ can largely
be understood from states near $E_{F}$, so we begin by analyzing the ground
state band and magnetic structure of the low-temperature (striped) phase of
CaFe$_{2}$As$_{2}$ within the LSDA.  Results for tetragonal and
orthorhombic striped structures are very similar, suggesting that the
slight difference in measured $a$ and $b$ lattice parameters plays a minor
role in the description of magnetic properties.  Experimental lattice
parameters were employed\cite{STR}.  CaFe$_{2}$As$_{2}$ has an internal
parameter $z$ which determines the Fe-As distance
$R_{\mathrm{{Fe}-{As}}}$.  Here we treat $z$ as a parameter whos main
effect is to control the magnetic moment $M$, which in turn strongly affects
the character of magnetic interaction, as we will show.

In Fig.~\ref{fig:dos}$(a)$, the density of states $N(E)$ is shown over a
2~eV energy window for a small and large moment case ($M$=0.4$\mu_{B}$ and
1.1$\mu_{B}$). Particularly of note is a sharp peak near $E_{F}$, of width
$\sim$50~meV.  This narrow band is found to consist almost entirely of
majority-spin Fe $d_{xy}$ and $d_{yz}$ orbitals.  The peak falls slightly
below $E_{F}$ for $M$=0.4$\mu_{B}$ and slightly above for $M$=1.1$\mu_{B}$.
Thus, as $R_{\mathrm{{Fe}-{As}}}$ is smoothly varied so that $M$ changes
continuously, this peak passes through $E_{F}$.  As a result $N(E_{F})$
reaching a maximum around $M$=0.8$\mu_{B}$ (Fig.~\ref{fig:dos}$(b)$).  This
point also coincides with the LSDA minimum-energy value of
$R_{\mathrm{{Fe}-{As}}}$.  That this unusual dependence originates in the
magnetic part of the Hamiltonian can be verified by repeating the
calculation in the nonmagnetic case.  As Fig.~\ref{fig:dos}$(b)$ shows, the
nonmagnetic $N(E_{F})$ is large, and approximately independent of
$R_{\mathrm{{Fe}-{As}}}$.  In summary, the magnetic splitting produces a
pseudogap in $N(E)$ for large $M$; the pseudogap shrinks as $M$ decreases
and causes a narrow band of Fe $d$ states to pass through $E_{F}$, creating
a sharp maximum in $N(E_{F})$ near $M$=0.8$\mu_{B}$. As we will show
below that moment demarcates a point of transition from itinerant to localized
behavior (see also Ref.~\cite{Sam}).

%--- chi0 ---

Next we turn to magnetic excitations.  In the standard picture of magnons in metals,
$\mathrm{{\operatorname{Im}}}\chi_{0}(\omega)$ is significant only for
frequencies exceeding the magnetic (Stoner) splitting of the $d$ bands,
$\epsilon_{d}^{\downarrow}${}$-${}$\epsilon_{d}^{\uparrow}$=$IM$ (\emph{cf}
Eq.~\ref{eq:jdos}). In such cases $\mathrm{{\operatorname{Im}}}\chi_{0}$ is
small at low $q$ and low energies and well defined magnons appear at energies near
$|1-I\mathrm{{\operatorname{Re}}\chi_{0}|=0}$ (\emph{cf} Eq.~\ref{eq:chi}). As
$\mathrm{{\operatorname{Im}}}\chi_{0}$ increases, it initially broadens the
(formerly sharp) SW spectrum $\bar\omega(\mathbf{q)}$; as it becomes large
the spectrum can become incoherent, or (Stoner) peaks can arise from
$\mathrm{{\operatorname{Im}}}\chi_{0}$, possibly enhanced
by small $1-I\mathrm{{\operatorname{Re}}}\chi_{0}$. Fig.~\ref{fig:dos}$(c)$
shows $\mathrm{{\operatorname{Im}}}\chi_{0}$($q$=0,\,$\omega$) on a
broad energy scale.  The picture we developed for $N(E)$ leads naturally to a
classification of Stoner transitions into three main types.

%\begin{enumerate}

%\item
(1) Excitations between the (largely) Fe $d_{xz}$ states centered near
$-$0.5~eV to $d$ states centered near 1~eV (for $M$=1.1$\mu_{B}$). The
magnetic splitting of these states matches well with the usual Stoner
splitting $IM$ in localized magnets, and scales with $M$. ($I${$\approx$}1~eV
in 3$d$ transition metals.) These high-energy transitions
%($\omega$ is too large to be detected by current neutron experiments)
are depicted by a large red arrow in Fig.~\ref{fig:dos}$(a)$ for
$M$=1.1$\mu_{B}$, and by vertical arrows in Fig.~\ref{fig:dos}$(c)$ showing
$\chi_{0}$(\emph{q}=0,$\omega$) for $M$=0.4, 0.8, and 1.1$\mu_{B}$.  They
are too high in energy to be observed by neutron measurements.

%\item
(2) Excitations from $d_{xz}$ to $d_{xy}$ and $d_{yz}$ states
just above the psueudogap, depicted by a small red arrow in
Fig.~\ref{fig:dos}({\emph{a}}), and slanting arrows in
Fig.~\ref{fig:dos}(\emph{d}).  These are the excitations probably detected
in neutron measurements \cite{AMESneut,ORNLneut}.  This pseudogap
is well defined for $M${}$\geq$1.1$\mu_{B}$, but is modified as
$M$ decreases, which leads to the following:

%\item
(3) Near $M$=0.8$\mu_{B}$, the narrow $d$ band passes through $E_{F}$,
opening up channels, not previously considered, for low-energy,
particle-hole transitions within this band.  When $M$ reaches 0.4$\mu_{B}$,
this band has mostly passed through $E_{F}$ and the pseudogap practically
disappears.

%.  Indeed, for M=0.4$\mu_{B}$ the pseudogap formation practically
%disappears and low energy spin-flip transitions are very intensive
%(Fig.1c). These excitations, which can be associated with Stoner
%transitions, are significant below 50 meV with their intensity strongly
%dependent on $\bfq$ and the size of $M$.

%\end{enumerate}

What makes the pnictide systems so unusual is that $\mathrm{{Im}}\chi_{0}$
is already large at very low energies ($\sim$10~meV) once the sharp peak in
$N(E)$ approaches $E_{F}$. One of our central findings is that this system
undergoes a \emph{ transition from localized to a coexistence of localized
and itinerant behavior} as $M$ decreases from $M${}$\gtrsim$1.1$\mu_{B}$ to
$M${}$\approx$0.8$\mu_{B}$.  Moreover, the itinerant character is of an
unusual type: elementary excitations are mostly single particle-hole like:
they can be well defined in energy and $q$.  Those represent
\emph{coherent} excitations.  The dependence of $N(E_F)$ on $M$ is not only
responsible for them, but also may explain the unusual linear
temperature-dependence of paramagnetic susceptibility, and the appearance
of a Lifshitz transition with Co doping \cite{BAFECO}.

%It also
%may be related to the observed anomalies in
%Ba(Fe$_{1-x}$Co$_{x}$)$_{2}$As$_{2}$ at low Co concentration [].

%--- full chi ---

%For chi(w) pictures see : raw-data-x0-x-ux0/new/
% For 392x-q100w.eps use raw-data-x0-x-ux0/new/plot.chix
% For swpAFM.eps use raw-data-x0-x-ux0/new/plot.sweAFM
% For chi100w0.9.eps use raw-data-x0-x-ux0/new/plot.chiwlargeq
\begin{figure}[ptbh]
\begin{tabular}[c]{cc}
\includegraphics[height=2.8cm]{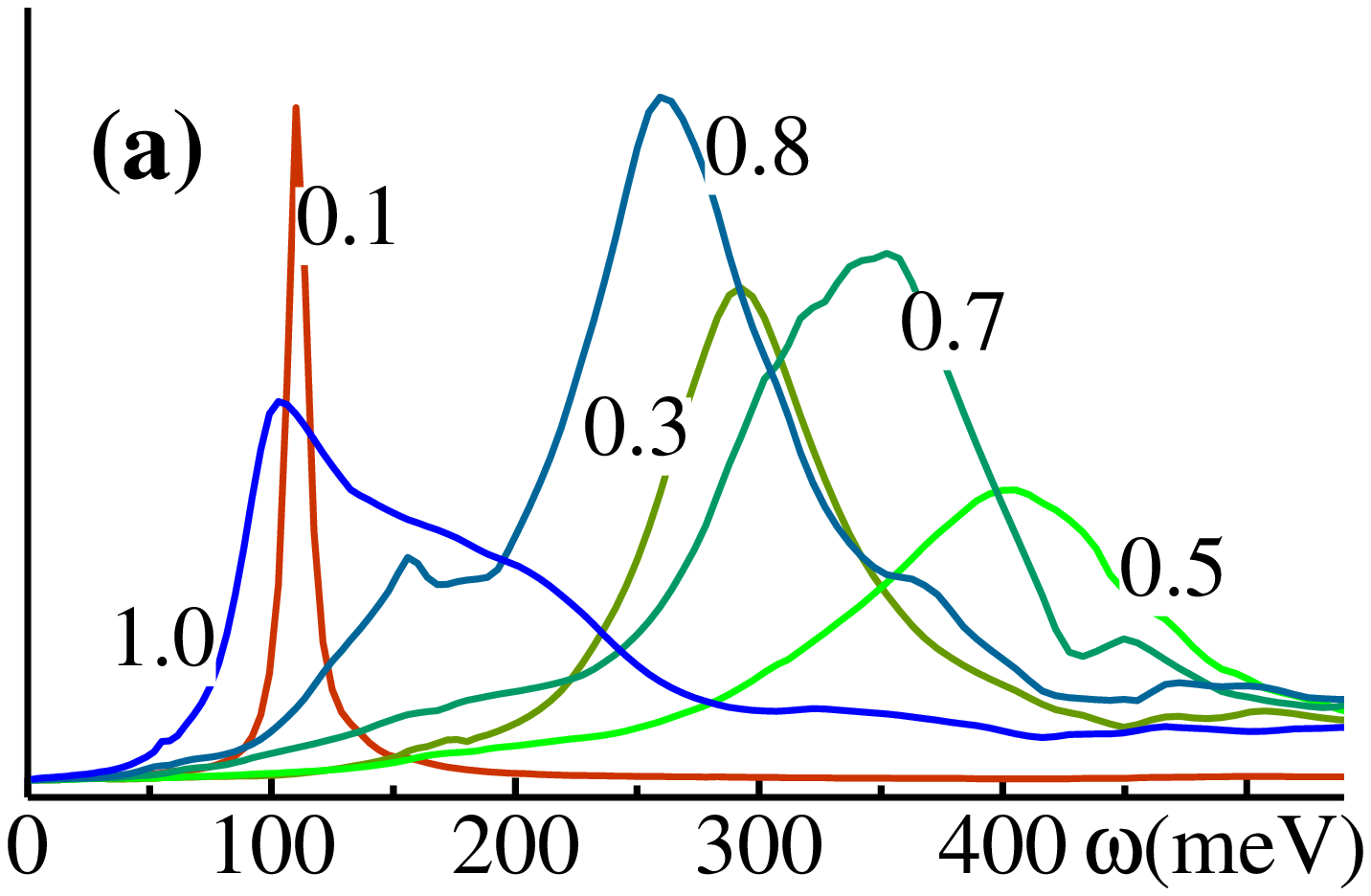} &
\includegraphics[height=2.8cm]{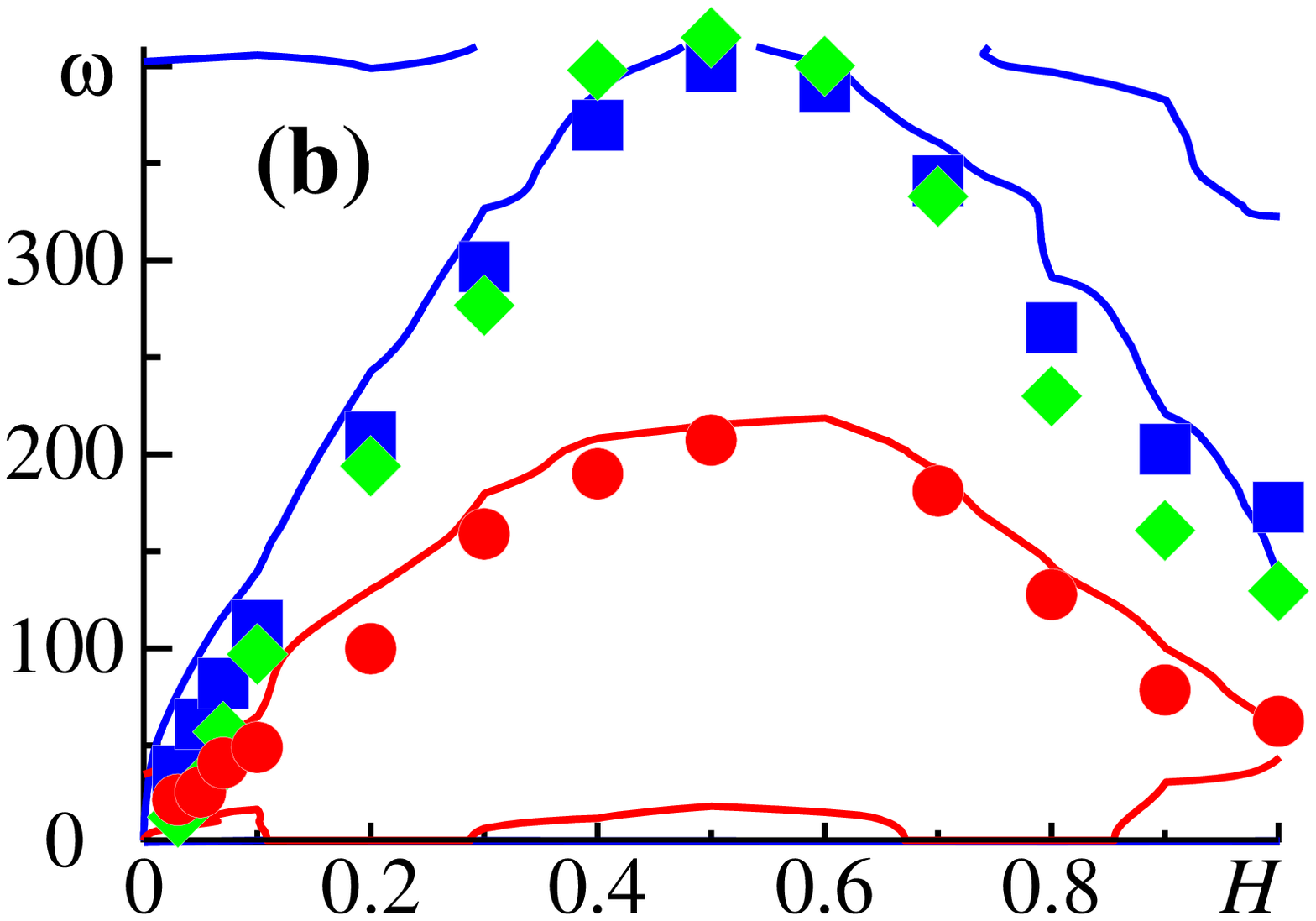}
\end{tabular}
\begin{tabular}[c]{cc}
\includegraphics[height=2.8cm]{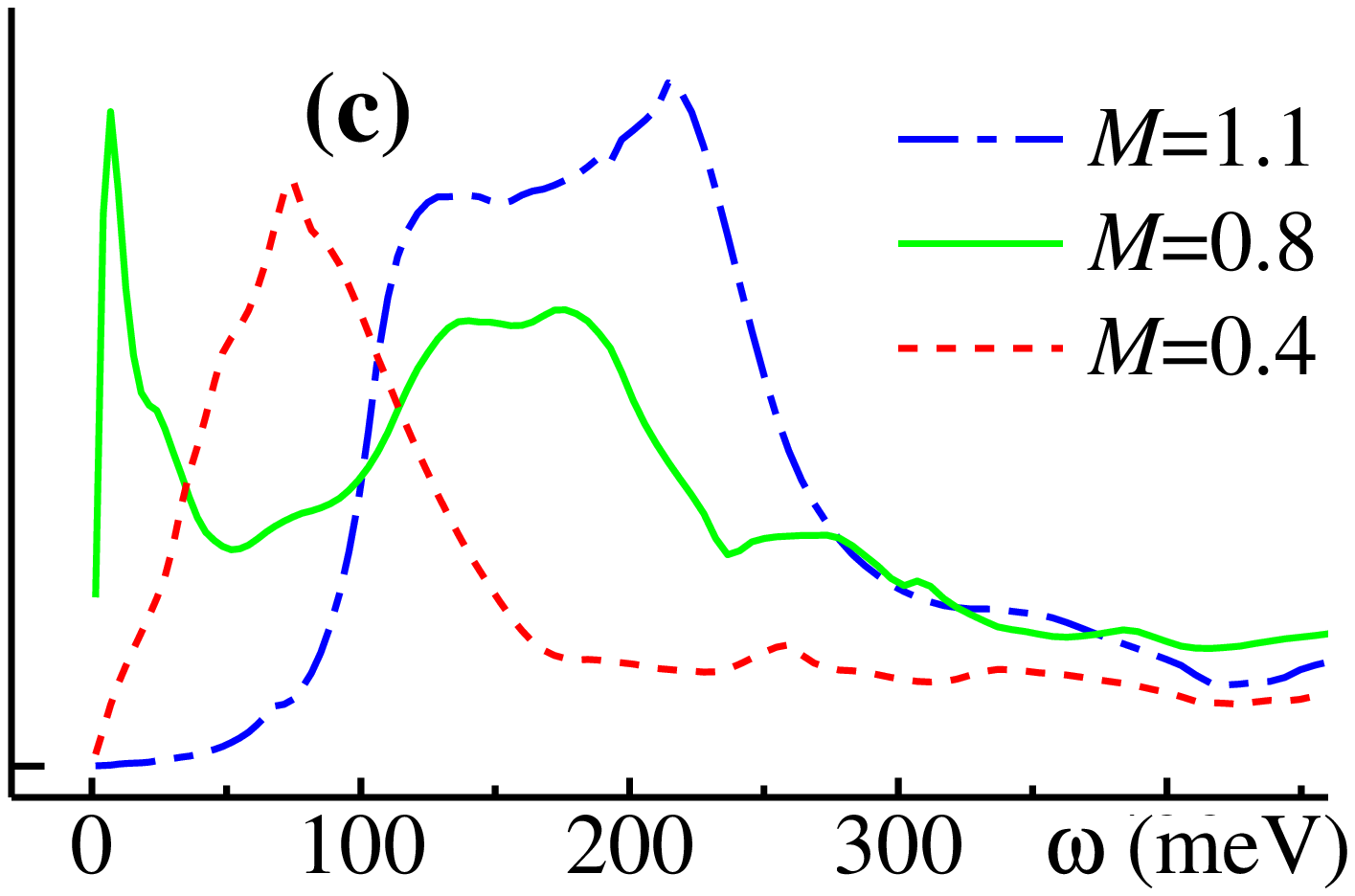} &
\includegraphics[height=2.8cm]{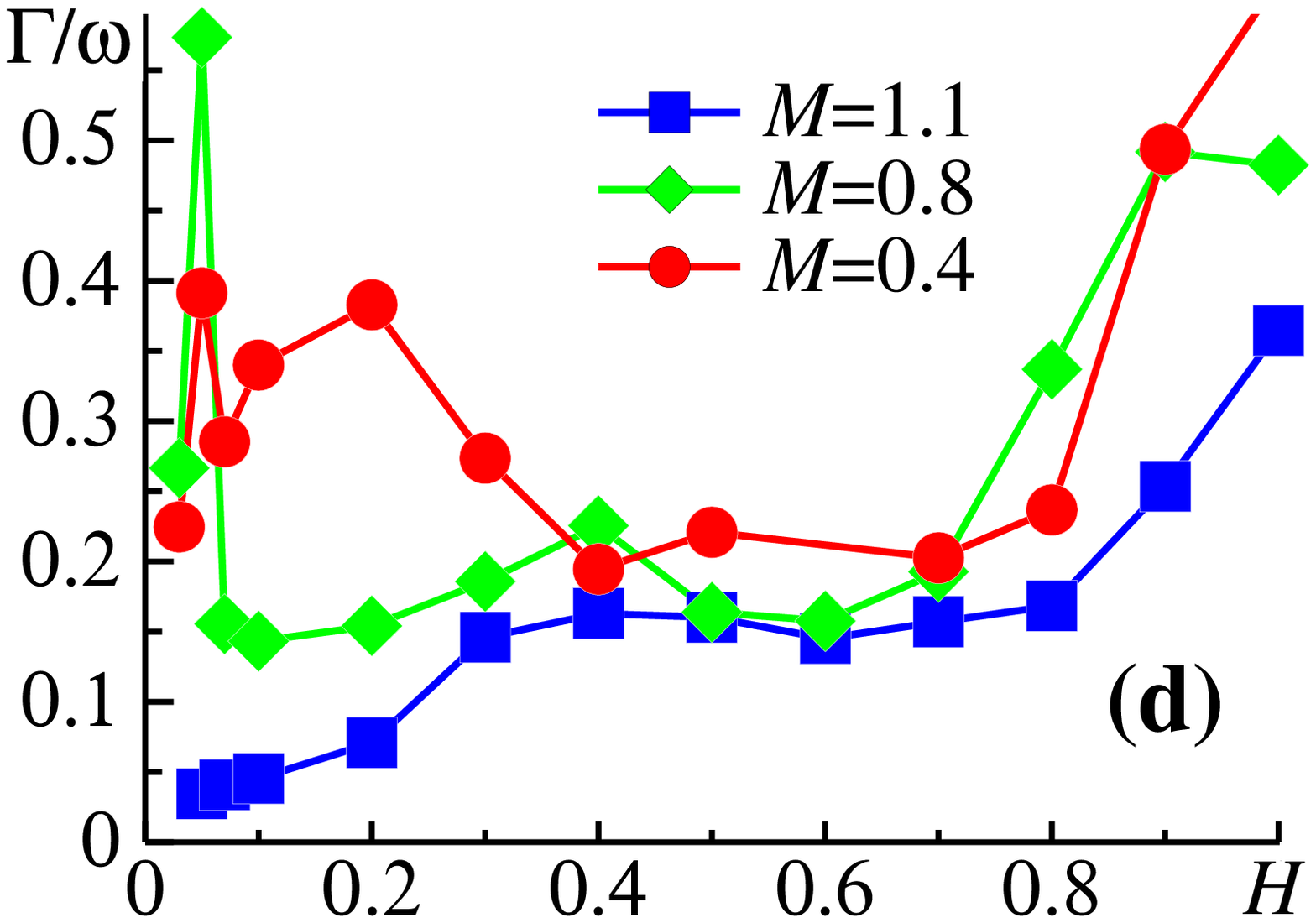}
\end{tabular}
\caption{$\mathrm{{Im}}\chi($\textbf{q}$,\omega)$ along the AFM axis,
$\mathbf{q}$=[$H$,0,0]$2\pi/a$. (\textbf{a}) $\mathrm{{Im}}\chi(\omega)$
for various $H$ and $M$=1.1$\mu_{B}$.  Extracted from
$\mathrm{{Im}}\chi(H,\omega)$ are peak positions $\bar\omega(H)$ (meV) and
HWHM $\Gamma$.  (\textbf{b}) and (\textbf{d}) depict
$\bar\omega$ and the ratio $\Gamma/\bar\omega$, for the three moments shown in the
key.  Panel (\textbf{b}) also depicts, as solid lines, contours
$|1-I\mathrm{{Re}\chi_{0}|=0}$ in the $(\omega,H)$ plane, for
$M$=1.1 and 0.4$\mu_{B}$.  Panel (\textbf{c}) shows
$\mathrm{{Im}}\chi$($H$=0.9,$\omega$) for the three moments.}
\label{fig:chi100}
\end{figure}

Fig.~\ref{fig:chi100} focuses on the AFM line,
$\mathbf{q}$=[$H$00]$2\pi/a$.  Panel (\textbf{a}) shows the full
$\mathrm{{Im}}\chi\left(\omega\right)$ for $M$=1.1$\mu_{B}$ for several
$q$-points spanning the entire line, 0{$<$}{$H$}{$<$}1.
%\cite{QREF}.
At low $q$, peaks $\bar\omega$ in $\chi$ are sharp; and $\bar\omega$
depends on $H$ in the expected manner ($\bar\omega${}$\propto${}$H$).
$\bar\omega$ reaches a maximum near $H$=1/2
(Fig.~\ref{fig:chi100}(\emph{b})), for all three values of $M$.  The peaks
broaden with increasing $q$; nevertheless we can associate them with
magnons, because they coincide closely with vanishing
$|1-I\mathrm{{Re}\chi_{0}|}$ and $\Gamma$ is not too large.  The magnon
character is preserved for most $q$ at all moments, as
Fig.~\ref{fig:chi100}(\emph{b}) shows: but for $H${$>$}0.8 the peaks are
strongly broadened, especially for small $M$.  $\bar\omega$ is in good
agreement with neutron data of Ref.\cite{ORNLneut}, except that neutron
data are apparently smaller than large $M$ calculations predict
\cite{DISCREPANCY}.

One experimental measure of the validity of the local-moment picture is the
ratio of half-width at half-maximum (HWHM) $\Gamma$ to $\bar\omega$, shown in
Fig.~\ref{fig:chi100}(\emph{d}). For large $M$ and most of $q$, 
$\Gamma/\bar\omega\sim$0.15-0.18. 
%This broadening occurs through Stoner excitations.
For intermediate and small $M$, $\Gamma/\bar\omega\geq$0.2 for a wide diapason
of $q$. This is significant: it reflects the increasing Stoner character of
the elementary excitations. Were there an abrupt transition into a
conventional Stoner continuum as argued in Ref.~\cite{ORNLneut}, it would
be marked by an abrupt change in $\Gamma/\bar\omega$.  This is not observed;
yet damping appears to increase with energy and $q$, reflecting normal
metallic behavior.

%We can elucidate this in more detail. Excitations for $H${$\lesssim$}0.2
%originate mostly from transitions within the cylindrical-like Fermi surface
%centered at $k$=(1/2,0,$k_{z}$). But for $H${$\rightarrow$}1, excitations
%originate mostly from transitions between different pockets in the Fermi
%surface, $\mathbf{q}$=(1/2,0,$k_{z}$){$-$}(0,0,$k_{z}$). For interpocket
%transitions, Im$\chi_{0}$ is large, which explains the increase in
%$\Gamma/\bar\omega$ as $H${$\rightarrow$}1. For intermediate and small $M$ the
%narrow $d$ band crosses $E_{F}$ and introduces a significant itineracy even
%for small $q$: Fig.~\ref{fig:chi100}(\emph{d}) shows clearly that
%$\Gamma/\bar\omega$ becomes small only for extremely small $q$.

Spectra for $H$=0.9 (Fig.~\ref{fig:chi100}(\emph{c})) adumbrate two
important findings of this work.  When $M$=0.8$\mu_{B}$, a sharp peak in
$\chi(\omega)$ appears near 10~meV.  There is a sharp peak in $\chi_0(\omega)$
at $\bar\omega${}$\approx$10~meV also, classifying this as a particle-hole
excitation originating from the narrow $d$ band depicted in
Fig.~\ref{fig:dos}.  Being well defined in energy it is coherent, analogous
to a SW, only with a much larger $\Gamma/\bar\omega$.  Yet it
is strongly enhanced by collective interactions, since
$|1-I\mathrm{{Re}\chi_{0}|}$ ranges between 0.1 and 0.2 for
$\omega${}$<$100~meV.  This new kind of itinerant excitation will be seen
at many values $q$, typically at small $q$.  The reader many note the sharp
rise and fall in $\Gamma/\bar\omega$ at small $q$ in
Fig.~\ref{fig:chi100}$(d)$.  This anomaly reflects a point where a SW and a
particle-hole excitation coalesce to the same $\bar\omega$.

Returning to $H$=0.9 when $M$=1.1, there is a standard (broadened) SW at
200~meV; \emph{cf} contours in Fig.~\ref{fig:chi100}(\emph{b}).  A
\emph{second}, low-energy excitation can be resolved near 120~meV.  This is
no corresponding zero in the denominator in Eq.~(\ref{eq:chi}), but no
strong peak in $\chi_0(\omega)$, either.  This excitation must be
classified as a hybrid intermediate between Stoner excitations and SWs.
Only a single peak remains when $M$=0.4$\mu_{B}$ and the peak in $N({E})$
has mostly passed through $E_{F}$ (Fig.~\ref{fig:dos}).

\begin{figure}[ptbh]
\begin{tabular}[c]{cc}
\includegraphics[height=2.8cm]{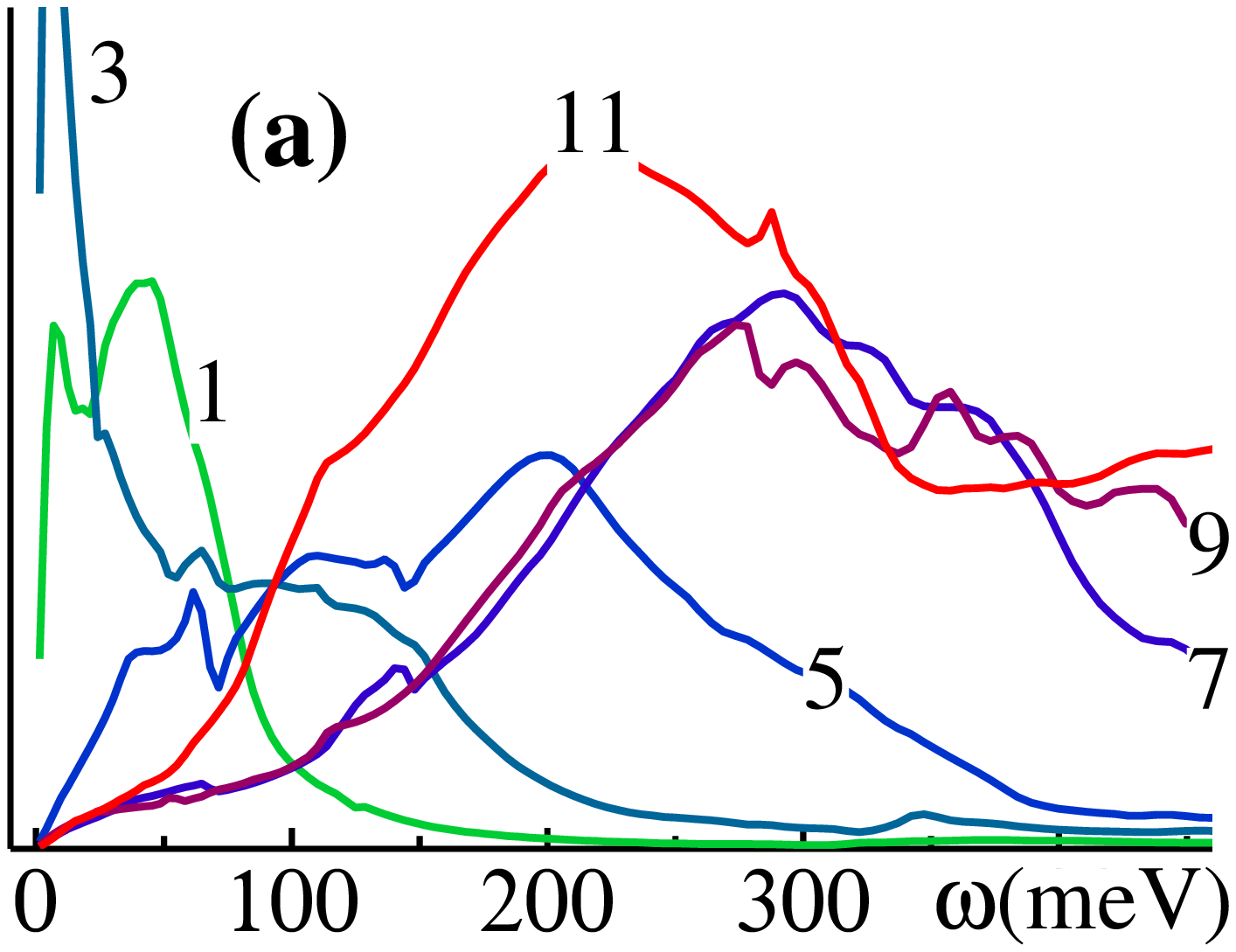} &
\includegraphics[height=2.8cm]{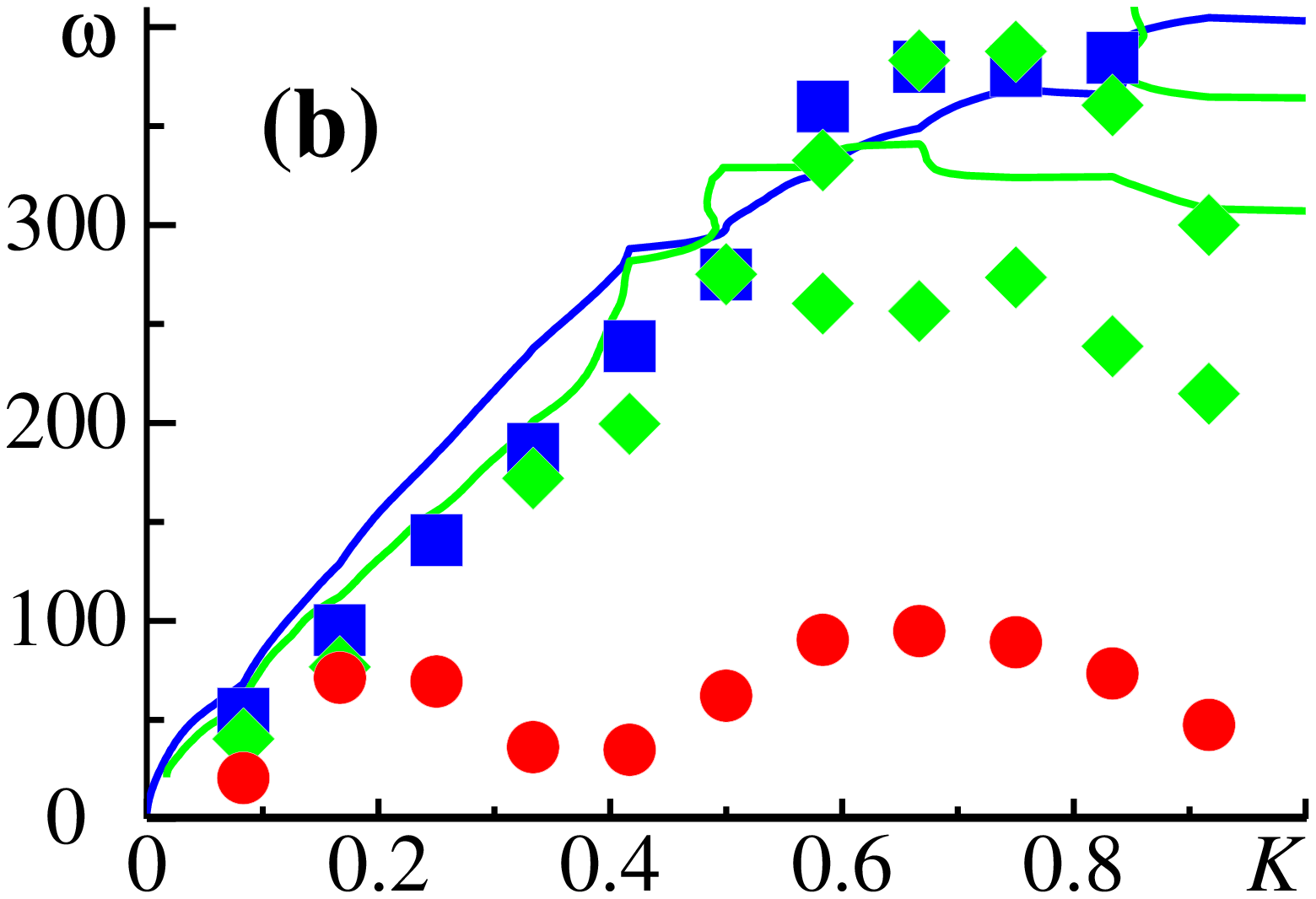}\\
\end{tabular} \\
\caption[]{${\rm{Im}}\chi($\bfq$,\omega)$ along the FM axis,
  $\bfq$=[0,$K$,0]$2\pi/a$.  ({\bf a}) ${\rm{Im}}\chi(\omega)$ at
  $K$= 1/12, 3/12, 5/12, 7/12, 9/12, 11/12, for $M$=0.8$\mu_{B}$.  Strong
  Stoner excitations can be seen for $K$=1/12 and 3/12 near
  $\bar\omega${$\sim$}10-20~meV.  ({\bf b}): Contours $|1-I\rm{Re}\chi_0|=0$ in
  the $(\omega,K)$ plane, for $M$=1.1$\mu_{B}$ and 0.8$\mu_{B}$, analogous
  to Fig.~\ref{fig:chi100}(\emph{b}), and dominant peak positions $\bar\omega$
  obtained by a nonlinear least-squares fit of one or two gaussian
  functions to $\chi(\omega)$ over the region where peaks occur.}
\label{fig:chi010}
\end{figure}

Along the FM line $\mathbf{q}$=(0,$K$,0)$2\pi/a$, $\chi(\omega)$
is more complex, and more difficult to interpret. At 
$M$=1.1$\mu_{B}$ sharp, well defined collective excitations are found
at low $q$, and broaden with increasing $q$.  There is a reasonably close
correspondence with the zeros of $|1-I\mathrm{{Re}\chi_{0}|}$ and peaks in
$\chi$, as Fig.~\ref{fig:chi010}(\emph{b}) shows.  The $M$=0.8$\mu_{B}$
case is roughly similar, except that for $K${$>$}0.4 excitations cannot be
described by a single peak.  
%Note also that zeros in
%$|1-I\mathrm{{Re}\chi_{0}|}$ occur at two distinct energies.  Roughly
%speaking $\bar\omega(K)$ remains monotonic (consider the average location of
%$|1-I\mathrm{{Re}\chi_{0}|}$=0), though the statement becomes a little
%ambiguous.
Note that for fixed $q$,
propagating spin fluctuations (characterized by peaks at $\bar\omega$) can
exist at \emph{multiple energies} --- the magnetic analog of the dielectric
function passing through zero and sustaining plasmons at multiple energies.
%They are the result of many-body interactions which lies outside the
%framework of the Heisenberg model.
Peaks in $\chi(\omega)$ can broaden as a consequence of this; the $K$=7/12,
9/12 and 11/12 data of Fig.~\ref{fig:chi010}(\emph{a}) are broadened in
part by this mechanism as distinct from the usual one (intermixing of
Stoner excitations).  Second, consider how $\partial\bar\omega/\partial{q}$
changes with $M$ for $K${$>$}0.5 (Fig.~\ref{fig:chi010}(\emph{b})).  When
$M$=1.1$\mu_{B}$, $\bar\omega$ increases monotonically with $K$.  For
$M$=0.8$\mu_{B}$, $\bar\omega$ has a complex structure but apparently
reaches a maximum before $K$ reaches 1.  That
$\partial\bar\omega/\partial{q}$ changes sign is significant: it marks the
disappearance of magnetic frustration between the ferromagnetically aligned
spins at low moments, and the emergence of stable FM order along [010],
characteristic of local-moment behavior (see also Ref.\cite{JIJUS}).
Experimentally, Ref.~\cite{ORNLneut} reports
$\partial\bar\omega/\partial{q}${$>$}0 for $K${$>$}0.5.  While our
calculations provide a clear physical interpretation, we note significant
differences in the moment where we observe this effect
($M$=0.8{-}1.1$\mu_{B}$) and the effective spin ($S$=0.2) used in
Ref.\cite{ORNLneut}.

\begin{figure}[ptbh]
\begin{tabular}[c]{cc}
\includegraphics[height=2.8cm]{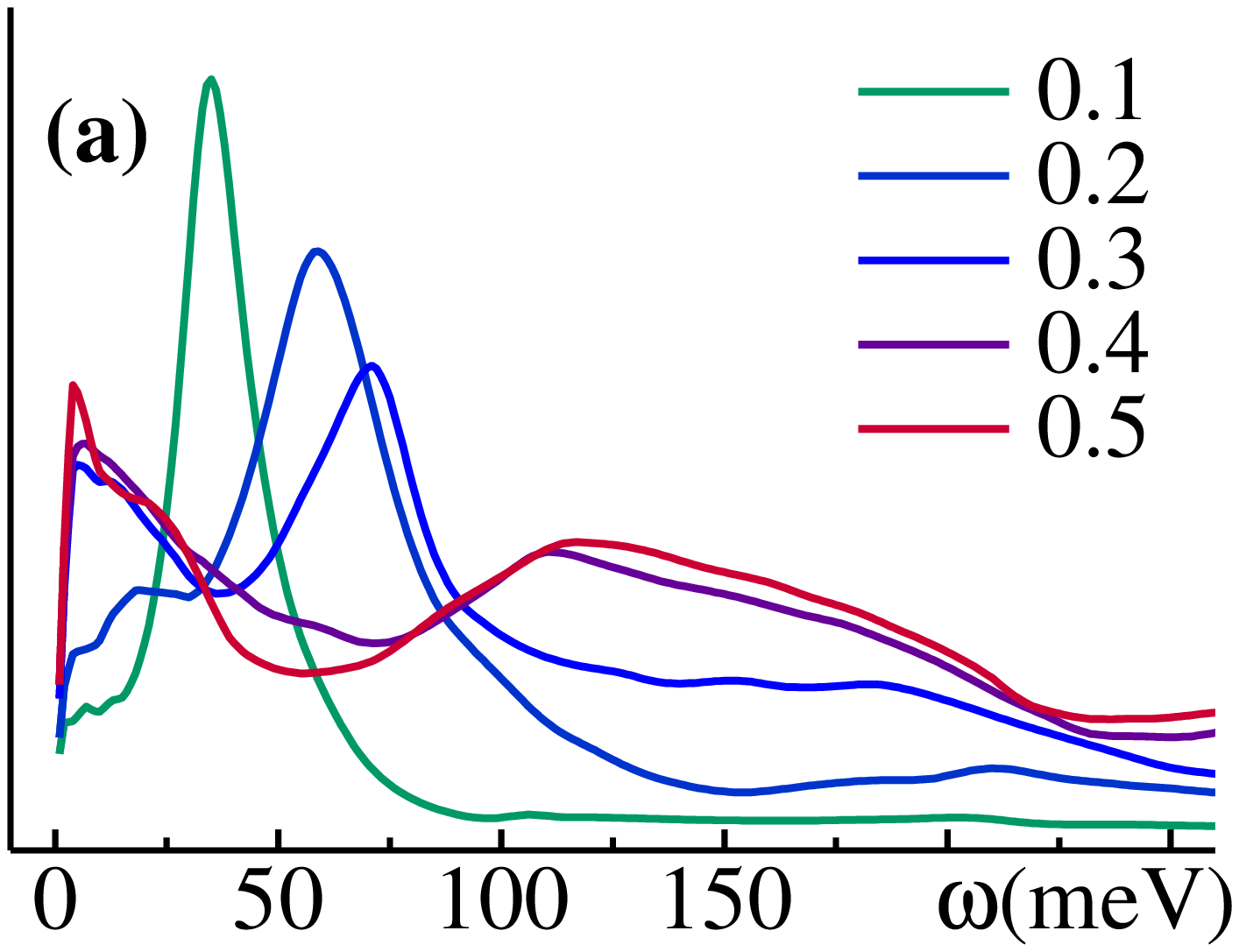}
\includegraphics[height=2.8cm]{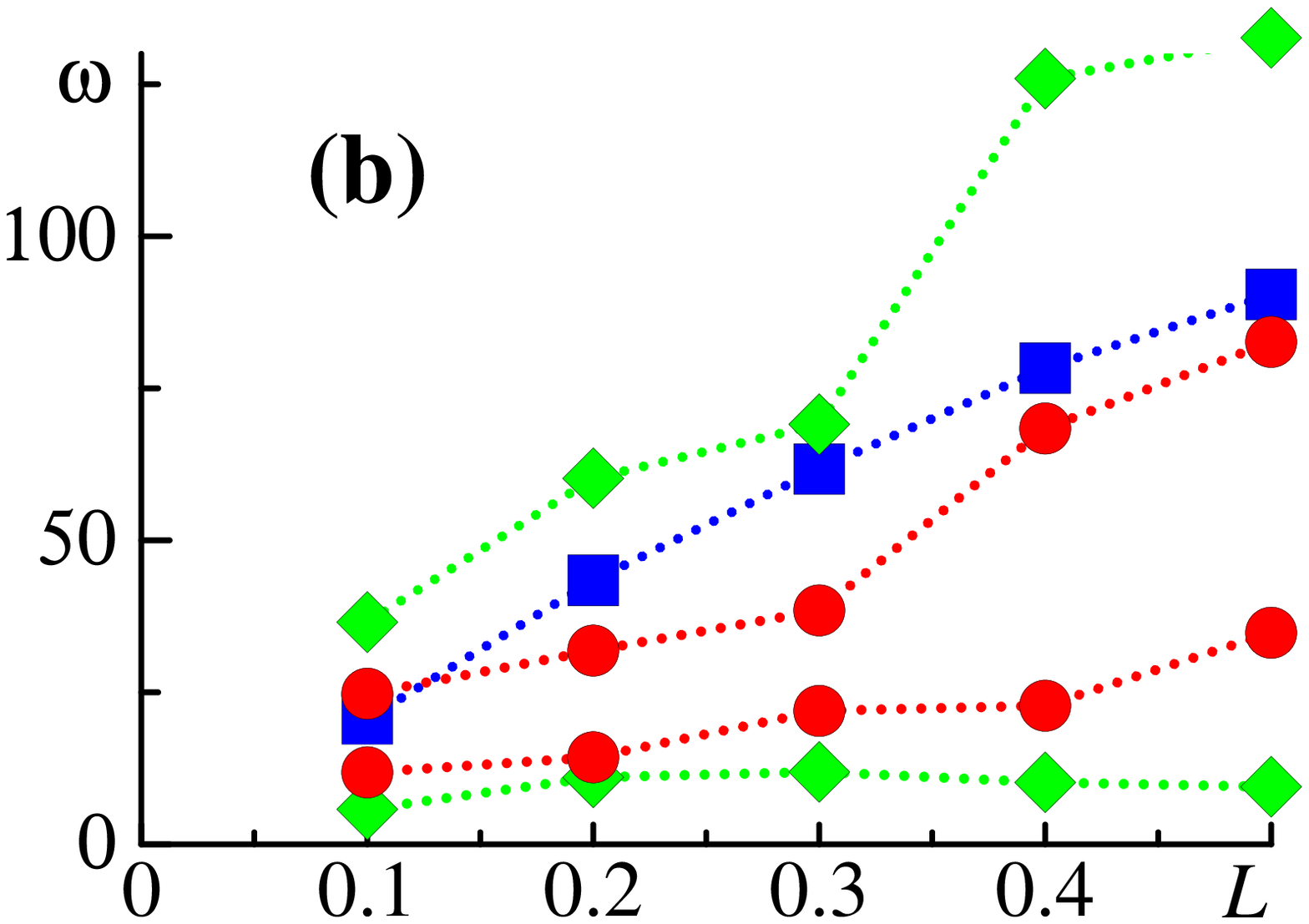}
\end{tabular}
%\begin{tabular}[c]{cc}
%\includegraphics[width=4.2cm]{doskefm10meV.eps}
%\includegraphics[width=4.2cm]{doskefp10meV.eps}
%\end{tabular}
\caption{$\mathrm{{Im}}\chi($\textbf{q},$\omega)$ along the $c$ axis,
$\mathbf{q}$=[0,0,$L$]$2\pi/c$. (\textbf{a})
$\mathrm{{Im}}\chi(\omega)$ at values of $L$ listed in the key,
for $M$=0.8$\mu_{B}$. When $M$=1.1$\mu_{B}$ (not shown) the spectrum is
well characterized by a single sharp peak at any $L$. When $M$=0.8 or 0.4$\mu_{B}$ 
the SW peak remains, but a low-energy Stoner excitation coexists with
it. The latter are most pronounced for larger $L$ but can be resolved at
every $L$.  Peak positions (single for $M$=1.1$\mu_{B}$, double for $M$=0.8 and
0.4$\mu_{B}$) are shown in (\textbf{b}).
%Right panels: Decomposition of $N(E_{F})$ into individual $k$ contributions when
%$N(E_{F})$ is maximum ($M$=0.8). Panels show scatter plots of
%$N$($E$,$\mathbf{k}$) looking down the (010) axis at two slightly different
%energies : $E$=$E_{F}$+10~meV (left) and $E$=$E_{F}${}$-$10~meV
%(right). Green arrows indicate the $\Gamma$ point, $k$=0. There is a
%roughly cylindrical Fermi surface centered at $k$=(1/2,0,$k_{z}$), and a
%smaller one centered at $k$=(0,0,$k_{z}$). Dot sizes and colors both
%indicate the magnitude of $N$($E$,$\mathbf{k}$), proportional to
%$|{\nabla_{k}} \epsilon_{k}|^{-1}$ on the constant energy (Fermi) surface at
%$E$: As $N$ ($E$,$\mathbf{k}$) increases, the color changes from blue
%($N$=100) to red ($N$=300). There are local ``hot spots'' in the
%(1/2,0,$k_{z}$) cylinder, whose locations are extremely sensitive to $E$,
%as comparison of the two figures show.
 }
\label{fig:chi001}
\end{figure}

Elementary excitations along the $c$ axis, $\mathbf{q}$=(0,0,$L$)$2\pi/c$,
bring into highest relief the transition from pure local-moment behavior
(collective excitations), to one where coherent itinerant Stoner and
collective excitations coexist.  Collective excitations are found for all
$L$ and all $M$.  For $M$=1.1$\mu_{B}$, a single peak is found: excitations
are well described by the Heisenberg model with weak
damping. Comparing Fig.~\ref{fig:chi001}(\emph{b}) to
Figs. ~\ref{fig:chi100}(\emph{b}) and ~\ref{fig:chi010}(\emph{b}), it is
apparent that $\bar\omega$ rises much more slowly along $L$ than along $H$
or $K$, confirming that interplane interactions are weak.  When $M$ drops
to 0.8$\mu_{B}$, a second-low energy peak at $\bar\omega^{\prime}$ emerges
at energies below 20~meV, for small $q$ along [0,$K$,0], and for all $q$
along [0,0,$L$], coexisting with the collective excitation.
%For $\bar\omega^{\prime}${$<$}20~meV and $M$=0.8$\mu_{B}$,
%$\bar\omega^{\prime}${$\propto$}$L$; for larger $L$ $\bar\omega^{\prime}$ becomes
%independent of $L$.  The peak is particularly pronounced as
%$L${$\rightarrow$}1/2; see Fig.~\ref{fig:chi001}(\emph{a}).
Why these transitions are absent for $M$=1.1$\mu_{B}$ and are so strong at
$M$=0.8$\mu_{B}$ can be understood in terms of roughly cylindrical Fermi
surface at $k$=(1/2,0,$k_{z}$).  Single spin-flip transitions between
occupied states $\epsilon_{\mathbf{k}}^{\uparrow}$ and an unoccupied states
$\epsilon_{\mathbf{k}+\mathbf{q}}^{\downarrow}$ separated by $\sim$10~meV
are responsible for this peak (see Eq.~\ref{eq:jdos}). They originate from
``hot spots'' where $N$($\epsilon_{\mathbf{k}}^{\uparrow}$)
and $N$($\epsilon_{\mathbf{k}+\mathbf{q}}^{\downarrow}$)
are both large.

%This only occurs for small $\mathbf{q}$ when
%$\mathbf{q}${}$\parallel$[010]; but when $\mathbf{q}${}$\parallel$[001];
%the entire cylinder at (1/2,0,$k_{z}$) can contribute: ``hot spots'' occur
%at points all along (1/2,0,$k_{z}$), but for any $k_{z}$ they appear over a
%very small energy window. 

%Thus the $k_{z}$ where a hot spot is found
%changes dramatically with small changes in $E$, as the last panels of
%Fig.~\ref{fig:chi001} show. A strong peak in $\chi$ appears when
%$\mathbf{q}$=$\mathbf{k}^{\downarrow}${}$-${}$\mathbf{k}^{\uparrow}${} is
%large for many values of $k$.

%There is a particularly large volume of the
%Brillouin zone in the vicinity of (1/2,0,1/2) where pairs of hot spots
%appear (Fig.~\ref{fig:chi001}, last panels); consequently the peaks are
%particularly pronounced as $L${$\rightarrow$}1/2.

From the \emph{SW velocities},
$(\partial\bar\omega$/$\partial q)_{q=0}$, anisotropy of the exchange couplings
can be determined \cite{JIJ,JIJUS,AMESneuta}.  We find strong
in-plane and out-of-plane anisotropies, and predict $v_{b}$/$v_{a}$=0.55 and
find $v_{c}$/$v_{a}$=0.35, where $v_{a}$=490meV$\cdot${\AA }.  Neutron
scattering experiments~\cite{AMESneuta,RevNeut} have measured with $v_{c}$/$v_{a-b}$
to be $\sim$0.2-0.5.

%The static calculations of
%susceptibility revealed anisotropy of the exchange coupling with smaller
%exchange coupling along FM line. Our calculations show that the effective
%interaction along this direction is fluctuating so the Heisenberg model
%with its static parameters may not be applicable. (The significant in-plane
%anisotropy of SWs spectrum has been obtained in Ref.~\cite{AMESneut},
%while no anisotropy has been detected in Ref.\cite{ORNLneut}.  The shape of
%SW spectrum along 010 direction confirms the negative sign (unstable
%FM order) for the exchange coupling along 010 direction found earlier in
%Refs.\cite{FESE,SAM,AMESneuta}.).

%Experimentally, the lowering of the magnetic moment and possible observation
%of these transitions can be done for Co doped 122 compounds\cite{BAFECO}.

In summary, we broadly confirm the experimental findings of
Refs.~\cite{AMESneut,ORNLneut}, that there is a spectrum of magnetic
excitations of the striped phase of CaFe$_{2}$As$_{2}$, which for the most
part are weakly damped at small $q$ and more strongly damped at large $q$.
A new kind of excitation was found, which originates from single
particle-hole transitions within a narrow band of states near $E_{F}$,
renormalized by a small denominator $|1-I\mathrm{{Re}\chi_{0}|}$
(Eq.~(\ref{eq:chi})).  They appear when the Fe moment falls below a
threshold, at which point the narrow band passes through $E_{F}$.  The
character of itineracy is novel: excitations occur at low-energy,
at energy scales typically below $T_{N}$, and can be sharply peaked, which
should make them accessible to experimental studies.  The distinction
between the two kinds of excitations is particularly observable in the
anomalous dependence of $\Gamma/\bar\omega$ on $q$ in
Fig.~\ref{fig:chi100}$(d)$.  Collective spin-wave-like excitations also are
unusual: multiple branches are found at some $q$. Finally, at higher $q$ we
find that the SW velocity changes sign, which is also observed
experimentally at similar values of $M$.

Overall a picture emerges where localized and itinerant magnetic carriers
coexist and influence each other.  This description falls well outside the
framework of a local-moments model such as the Heisenberg hamiltonian.  The
low-energy particle-hole excitations are strongly affected by Fe-As
distance (and thus by lattice vibrations) while the SWs are much less so.
While phonons and Stoner excitations are generally considered separately,
these findings suggest that they strongly influence each other.

%Some aspects of our prediction can be confirmed
%by new experiments, such as the frequency-dependent electrical conductivity
%as a function of doping or pressure.

This work was supported by ONR, grant N00014-07-1-0479 and by DOE contract
DE-FG02-06ER46302.  Work at the Ames Laboratory was supported by DOE
Basic Energy Sciences, Contract No. DE-AC02-07CH11358.


\begin{thebibliography}{99}
\bibitem {DISCREPANCY}The origin of the discrepancy in $\bar\omega$ with
experiment is not clear. Our calculation contains several important
approximations (local potential, linear response, rigid spin).
\end{thebibliography}
\end{document}